\documentclass[twocolumn,showpacs,preprintnumbers,amsmath,amssymb]{revtex4}
\usepackage{graphicx}% Include figure files
\usepackage{dcolumn}% Align table columns on decimal point
\usepackage{bm}% bold math

\begin{document}

\preprint{APS/123-QED}

\title{Weakly Nonlocal Hydrodynamics and The Origin of Viscosity in the 
Adhesion Model}% Force line breaks with \\

\author{A.L.B. Ribeiro}
\email{albr@uesc.br}
\author{J.G. Peixoto de Faria}
\email{jgfaria@uesc.br}
 \affiliation{Laborat\'orio de Astrof\'{\i}sica Te\'orica e Observacional\\
Departamento de Ci\^encias Exatas e Tecnol\'ogicas\\
Universidade Estadual de Santa Cruz}
%\affiliation{%

%\affiliation{

\date{\today}% It is always \today, today,

\begin{abstract}
We use the weakly nonlocal hydrodynamics approach
to obtain a dynamical equation for the peculiar velocity field
in which the viscosity term is physically motivated. Based on some
properties of the Ginzburg-Landau equation and the wave mechanics
analog of hydrodynamics we find the nonlocal adhesion approximation
taking into account the internal structures of the 
Zeldovich pancakes. If the internal structures correspond to
significant mesoscopic fluctuations, viscosity is probably
driven by a stochastic force and dynamics is given by the
noisy Burgers equation.
\end{abstract}


\maketitle

~~~~~
\vspace{0.5 truecm}
\section{\label{sec:intro}Introduction}

The complexity of the nonlinear structure formation process
is not possible to be studied just using  analytical methods.
But the comparison of analytical approximations and N-body experiments
may help us to understand the physics of highly nonlinear collective
phenomena. The Zeldovich
approximation was the first analytical try to predict large-scale
structures like pancakes, filaments and voids \cite{1}.
The problem with this kinematical  model is that particles pass
through the pancakes instead of
clustering  into smaller objects like
groups and galaxies. An important
improvement of the Zeldovich model is the
adhesion approximation \cite{2}, in which
a viscous term is introduced 
to prevent particle crossing. However, this
modification is not intended to describe the
internal structure of pancakes and, then, its
form has no constraint. Usually, a simple {\it ad-hoc}
term, $\nu\nabla^2{\bf u}$, is assumed and hence
viscosity is considered as an artificial effect in the large-scale fluid,
despite the 
good agreement of the model with numerical simulations (e.g. \cite{3}).
Actually, there are some alternative approaches (without introducing viscosity)
to the nonlinear regime like
the linear potential and the frozen flow approximations 
(e.g. \cite{4},\cite{5}). But a systematic comparison of several
statistics for the distributions resulting from these
approximations shows that the adhesion model works better 
over the nonlinear regime \cite{6}. This situation suggests 
the need of a more physical motivation to invoke the viscous term.
In this work, we use weakly nonlocal hydrodynamics to obtain a dynamical 
equation for the large-scale fluid in which 
the viscosity term is physically motivated.

\section{Hydrodynamics and Adhesion Approximation}

Let us define the peculiar velocity field as 
the vector field ${\bf u}=\dot{{\bf r}}-H{\bf r}$
(where $H=\dot{a}/a$ and
$a$ is the scale factor).
Consider the fluid motion near a point $0$. Then

\begin{equation}
u_i({\bf r})=u_i(0)+r_j{\partial u_i\over \partial r_j}\Bigg|_0 
+{1\over 2}r_jr_k{\partial^2u_i\over\partial r_jr_k}\Bigg|_0 +...
\end{equation}

\noindent Thus,

\begin{equation}
u_i({\bf r})-u_i(0)\approx r_j{\partial u_i\over \partial r_j}\Bigg|_0.
\end{equation}

\noindent The quantity ${\partial u_i\over \partial r_j}\equiv 
(\nabla{\bf u})_{ij}$ is called the velocity gradient tensor. The symmetric
part of the velocity gradient is the rate of the strain tensor

\begin{equation}
e_{ij}={1\over 2}\left({\partial u_i\over\partial r_j} + 
{\partial u_j\over \partial r_i}\right)
\end{equation}

\noindent and the antisymmetric part is the vorticity tensor

\begin{equation}
\Omega_{ij}={1\over 2}\left({\partial u_i\over \partial r_j} -
{\partial u_j\over\partial r_i}\right)=-{1\over 2}\epsilon_{ijk}\omega_k,
\end{equation}

\noindent where the vector ${\boldsymbol{ \omega}}$ is the vorticity 
${\boldsymbol {\omega}}=\nabla\times 
{\bf u}$. Thus

\begin{equation}
u_i({\bf r})-u_i(0)\approx r_je_{ij} + {1\over 2}({\boldsymbol {\omega}}\times
{\bf r})_i.
\end{equation}

\noindent We assume an irrotational
large-scale astrophysical
fluid,  which is  modelled as an incompressible
Newtonian fluid with constitutive equation given by

\begin{equation}
\sigma_{ij}=-p\delta_{ij}+\sigma^\prime_{ij},
\end{equation}

\noindent where $\sigma_{ij}$ is the stress tensor,  $p$ is
the thermal pressure and $\sigma^\prime_{ij}$ is the
shear stress tensor. Assuming that the shear tensor is linearly
related to the  velocity gradient tensor, we have, for an isotropic fluid

\begin{equation}
\sigma^\prime_{ij}=A_{ijkl}{\partial u_k\over\partial r_l},
\end{equation}

\noindent with $A_{ijkl}=\lambda\delta_{ij}\delta_{kl} +
\mu\delta_{ik}\delta_{il} + \mu^\prime\delta_{il}\delta_{jk}$
and $\mu=\mu^\prime$ for a symmetric tensor. Thus, the consitutive
equation becomes

\begin{equation}
\sigma_{ij}=-p\delta_{ij}+2\mu e_{ij}.
\end{equation}

\noindent Introducing this into the
momentum equation

\begin{equation}
\rho{Du_i\over Dt}=-\rho\nabla\phi + {\partial \sigma_{ij}\over \partial r_j}
\end{equation}

\noindent  and making 
the spatial derivatives with respect
to ${\bf x}={\bf r}/a(t)$, we find the Euler equation
for the peculiar velocity field

\begin{equation}
{\partial{\bf u}\over \partial t}+ {1\over a}({\bf u}\cdot\nabla){\bf u}
=  -{1\over a}\nabla\phi -{\dot{a}\over a}{\bf u} 
+ {\mu\over\rho a^2}\nabla^2{\bf u}
\end{equation}

\noindent where $\mu/\rho\equiv\nu$ is the kinematic viscosity,
$\phi$ is the gravitational potential and,
for the scales of interest, we have neglected the thermal pressure.

Now, changing the time variable from $t$ to $b$ (the growing mode of
the linear theory) and rescaling the peculiar velocity
and the potential gravitational as

\begin{equation}
{\bf v}={{\bf u}\over a\dot{b}}={{\rm d}{\bf x}\over {\rm d}b}
~~~{\rm and}~~~\varphi=\left({3\over 2}\Omega_0\dot{a}^2b\right)^{-1}\phi
\end{equation}

\noindent we find

\begin{equation}
{\partial {\bf v}\over\partial b} + ({\bf v}\cdot\nabla){\bf v}=
{3\over 2}{\Omega_0\over bf^2}\nabla(\theta -\varphi)+\nu\nabla^2{\bf v}.
\end{equation}

\noindent \noindent where $f(t)={{\rm d}\ln{b}\over {\rm d}\ln{a}}$,
$\Omega_0=8\pi G\rho_0/3H_0^2$ and we have taken
the velocity field  as the gradient of a velocity
potential (${\bf v}\equiv\nabla\theta$) 
as long as the motion is linear and streams of
particles do not cross.

 Assuming that over the mildly nonlinear regime the
gravitational potential is approximately equal to the velocity
potential, we finally get

\begin{equation}
{\partial {\bf v}\over\partial b} + ({\bf v}\cdot\nabla){\bf v}=
\nu\nabla^2{\bf v}
\end{equation}

\noindent which is known as the adhesion approximation
\cite{2}. For small $\nu$ it reduces to the Zeldovich model
which corresponds to a simple inertial motion. 
The viscosity term in the right side of (13) is introduced just
to prevent the inertial broadening of the first nonlinear
structures -- the Zeldovich pancakes. Hence, it is not a real viscosity
, in the sense of a significant velocity
gradient in the fluid, but a term which forces particles to stick
together. 

Despite the use of  a  mock
viscosity, the adhesion model shows a remarkable agreement
with the large-scale structure
produced in N-body simulations \cite{3,6}. Thus, any simple
improvement to
this model does not need to change  its  mathematical
structure very much, but just find a less artificial way to introduce the term in the 
right side of (13). Particularly, in this work, 
we want to find a physically motivated ``viscosity'' term presenting
a dependence on the internal
structure of the pancakes.

\section{Weakly Nonlocal Hydrodynamics}

The correct
framework to study continuum physics dealing with internal structures
is the weakly nonlocal (coarse grained) 
hydrodynamics \cite{7}. In this context, we can
use the Ginzburg-Landau equation as a first weakly nonlocal extension
of a homogeneous relaxation equation of an internal variable.
The Ginzburg-Landau equation is a modulational equation for
a complex function, $\psi$, which describes
weakly nonlinear phenomena in continuous media with linear dispersion
of general type. It is given by

\begin{equation}
{\partial\psi\over \partial t}=\psi + (1+ic)\nabla^2\psi
- (1+id)|\psi|^2\psi
\end{equation}

\noindent where the parameters $c$ and $d$ are the linear and
nonlinear dispersions. Some important properties of this equation
deserve special attention.

i) It can be reduced to a dissipative extension of the
nonlinear Schr\"odinger equation \cite{8}

\begin{equation}
i{\partial\psi\over\partial t}=\nabla^2\psi\pm V(\psi)\psi.
\end{equation}

ii) It admits amplitude-phase representation in the form \cite{9}

\begin{equation}
\psi=R\exp{i\zeta}.
\end{equation}

iii) It can be derived from a variational 
principle. In particular, weakly nonlocal fluids respecting
the second law of thermodynamics depends on the nonlocal potential
given by 

\begin{equation}
U=\delta\int\rho s(\rho,\nabla\rho)\;dV
\end{equation}

\noindent where $s$ is the entropy density \cite{10}.

iv) An important case, which allows us to
connect hydrodynamics and wave mechanics with a rich
structure of stationary solutions, is the 
Schr\"odinger-Madelung potential energy \cite{10}

\begin{equation}
U=-{\hbar^2\over 2m}{\nabla^2 R\over R}
\end{equation}

\section{Nonlocal Adhesion Model}

The wave mechanics analog of classical hydrodynamics is a well
known subject in the literature (see e.g. \cite{11}, \cite{12}),
but the use of this equivalence as an approach to study structure
formation in the universe is relatively new (e.g. \cite{13}, \cite{14}).
The basic idea is very simple: to describe collisionless matter as 
a classical field $\psi(x,t)$ obeying the Schr\"odinger and Poisson
equations

\begin{equation}
i\hbar{\partial\psi\over\partial t} = \left( -{\hbar^2\over 2m}\nabla^2 + V\right)\psi,
~~~ \nabla^2\phi=4\pi G\psi\psi^\ast;
\end{equation}

\noindent 
where $V=m\phi$ and $\hbar$ is a parameter
controlling the spatial resolution $\lambda$, according to the de Broglie
relation $\lambda=\hbar/mv$. This approach has the advantage of
working only with three spatial coordinates (plus time) and following
multiple streams in the phase space.
Promising results of this technique
were found for the collapse 
of a self-gravitating object and to 
simulate a two-dimensional Cold Dark Matter universe \cite{13}. 

Here, we make the wave analog of a nonlocal fluid under
a smooth gravitational field plus a coarse grained field by
introducing the amplitude-phase ansatz 
$\psi = R \exp{({i\over \hbar}S)}$ \cite{15} into 
the Schr\"odinger equation (19).
Equating
real and imaginary parts separetely, we find two coupled equations
for the classical action $S$ and the real amplitude $R$:

\begin{equation}
{\partial S\over\partial t} +{1\over 2m}(\nabla S)^2 + V +U=0
\end{equation}

\begin{equation}
{\partial R\over\partial t}= -{1\over m}\nabla R\cdot \nabla S - {1\over 2m}
R\nabla^2S,
\end{equation}

\noindent where $U$ is the potential previously defined in (18).
Note that equation (21) can be rewritten as

\begin{equation}
{\partial R^2\over\partial t} = -{1\over m}\nabla (R^2\nabla S).
\end{equation}

\noindent But 
${\bf p}=\nabla S$ in the Hamilton-Jacobi canonical
transformation. It then follows that ${\bf u}=\nabla S/m$ 
and (22)
takes the form of the hydrodynamical continuity equation

\begin{equation}
{\partial \rho\over \partial t} + \nabla\cdot(\rho{\bf u})=0.
\end{equation}

\noindent We can therefore attribute to the square of the amplitude
of the wave function the interpretation of a fluid density  whose
flux is conserved over time:
$\rho=\psi^\ast\psi=|\psi|^2$. Then, we finally get

\begin{equation}
\psi({\bf x},t)=\sqrt{\rho({\bf x},t)}\exp{[iS/\hbar]}.
\end{equation}

\noindent with $\rho=R^2$.

Now, note that Equation (20) can be rewritten as

\begin{equation}
m{\partial{\bf u}\over \partial t} + m({\bf u}\cdot\nabla){\bf u} +
\nabla(V + U)=0
\end{equation}

\noindent or

\begin{equation}
{\partial{\bf u}\over \partial t} + ({\bf u}\cdot\nabla){\bf u} +
\nabla(\phi + \phi_{vis})=0
\end{equation}

\noindent where the quantity

\begin{equation}
\phi_{vis}={U\over m}=-{\nu^2\over 2}\left[{\nabla^2\rho\over\rho} - 
{1\over 2}\left(
{\nabla\rho\over \rho}\right)^2\right]
\end{equation}

\noindent is the ``viscosity'' (or coarse grained)
 potential. Note that $\hbar/m$ has the dimensions of a
kinematical viscosity, such that we have defined $\nu\equiv\hbar/m$.

Using the same variable
transformation we made in  Section II for an expanding
background, we finally come to

\begin{equation}
{\partial {\bf v}\over\partial b} + ({\bf v}\cdot\nabla){\bf v}=
{3\over 2b}\nabla(\theta -\varphi -\phi_{vis})
\end{equation}

\noindent Again, for the weakly nonlinear regime, we have 
$\varphi\approx\theta$,
and then

\begin{equation}
{\partial {\bf v}\over\partial b} + ({\bf v}\cdot\nabla){\bf v}=
-{3\over 2b}\nabla\phi_{vis}
\end{equation}

\noindent which is the weakly nonlocal adhesion approximation. It is important
to note that the origin of viscosity should be related to
the divergence of some  ``shear'' tensor. Indeed, if we define

\begin{equation}
\sigma_{ij}^{kin}\equiv {\nu^2\over 4}\left[{\nabla_i\nabla_j\rho\over \rho} - 
{(\nabla_i\rho)(\nabla_j\rho)\over \rho}\right]={\nu^2\over 4}\nabla_i\nabla_j
\ln{\rho}
\end{equation}

\noindent we have that the divergence of this tensor is equal
to the density force associated with $\phi_{vis}$

\begin{equation}
\nabla\cdot\sigma=-\nabla\phi_{vis}.
\end{equation}

\noindent Now, defining a  ``kinematical velocity'' as

\begin{equation}
{\bf u}_{kin}\equiv{\nu\over 2}\nabla\ln{\rho},
\end{equation}

\noindent the symmetric part of the velocity gradient 
$(\nabla{\bf u}_{kin})_{ij}$ is

\begin{equation}
e_{ij}^{kin}=\rho{\nu\over 4}\left({\partial {(u_{kin})}_i\over\partial r_j} + 
{\partial {(u_{kin})}_j\over \partial r_i}\right)
\end{equation}

\noindent and 

\begin{equation}
\sigma_{ij}^{kin}=\rho\nu\left({\partial {(u_{kin})}_i\over\partial r_j} + 
{\partial {(u_{kin})}_j\over \partial r_i}\right)
\end{equation}

\noindent thus showing the connection between coarse grained viscosity
and nonlocal phenomena through the kinematical shear tensor.
The kinematical velocity was first introduced by \cite{11} and
it is not related to the organized mechanical motion,
but to the perturbations due to the coarse grained nature
of the fluid. From (34),  we have

\begin{equation}
{\partial {\bf v}\over\partial b} + ({\bf v}\cdot\nabla){\bf v}=
\nu\nabla^2{\bf v}_{kin}.
\end{equation}

\noindent The kinematical velocity introduces a deviation with respect
to the field ${\bf v}$. If we define
$\Delta{\bf v}_{kin}\equiv {\bf v} - {\bf v}_{kin}$, 
we find the general expression

\begin{equation}
{\partial {\bf v}\over\partial b} + ({\bf v}\cdot\nabla){\bf v}=
\nu\nabla^2{\bf v} + \nabla\eta
\end{equation}

\noindent where $\nabla\eta\equiv-\nu\nabla^2\langle\Delta{\bf
  v}_{kin}\rangle$ 
is a noise term. 
Hence, the weakly nonlocal adhesion model corresponds to the noisy Burgers
  equation.
Note that we recover the Zeldovich approximation  if 
${\bf v}_{kin}\rightarrow 0$, which points out the importance of internal structures to
prevent inertial broadening.

\section{Discussion}

The wave-mechanical analog of hydrodynamics
seems to be a promising framework
to find alternative solutions to both analytical and numerical
problems over the nonlinear regime of structure formation.
In this work, we have investigated the origin of the
viscosity term in the adhesion  model. 
Viscosity is the property which
makes particles stick together once they enter 
into the caustics predicted by the Zeldovich approximation,
just mimicking gravitational effects on smaller scales.
This approximation  describes the general nonlinear
structures but does not trace the internal dynamics of the pancakes. 
In this work, we assume the nonlocal hydrodynamics
description in the amplitude-phase representation to introduce
a physically motivated ``viscosity'' term into the adhesion model.
Using the Schr\"odinger-Madelung potential, we come to
a new equation for the adhesion model as a function
of a viscosity potential, which can be written as the divergence of
a new shear tensor. 

Our model identifies the origin of viscosity as a nonlocal effect.
Notice that the fluid equations are only
valid  when the average of the microscopic linear momentum equals the
product of the averages of density and velocity. However, due to the
mesoscopic nature of the averaging procedure, this is not
guaranteed. Generally, local fluctuations in the fluid are small enough
to fade away over averaging. But nonlocal (configurational) noise
can add on the hydrodynamics noise, leading to non-selfaveraging 
properties in the matter flows. Mesoscopic arrangements
in the system are sources of fluctuations, since  the averaging
procedure in the free volumes around them can lead to different results.
This effect is a source of nonlocal noise \cite{16}.
In the case of pancakes, the internal arrangements could be
the spatial configuration of protogalactic gas clouds
developed over the path followed
by matter (inside pancakes) to form filaments and knots.
These intermediate structures can be considered as
sources of the noisy mesoscopic fluctuations. 

Applications of (36)  to the nonlinear regime and comparisons to
N-body simulations
are beyond the scope of this paper and will be presented in a future
contribution. For the moment, we conclude that
the origin of viscosity in the adhesion model is probably associated
with weakly nonlocal effects in the internal structure of the pancakes. These
effects are given by some viscosity potential. In the case of
significant mesoscopic fluctuations, we assume
that the potential has a stochastic nature and the
dynamics is given by the noisy Burgers equation.

\begin{acknowledgments}
This work has the finantial support of CNPq
(grants 470185/2003-1) and UESC
(grants 220.1300.357 and 220.1300.324).
\end{acknowledgments}

%\newpage 


\begin{thebibliography}{}% Produces the bibliography via BibTeX.
\bibitem{1} Zeldovich, Y.B. 1970, A\&A {\bf 5}, 84.
\bibitem{2} Gurbatov, S.N., Saichev, A.L. and Shandarin, S. 1989, 
MNRAS {\bf 236}, 385.
\bibitem{3} Koffman, L., Pogosyan, S.  and Shandarin,  S. 1990, MNRAS
{\bf 242}, 200.
\bibitem{4} Bagla, J.S. and Padmanabhan, T. 1993, ApJ {\bf 417}, 3.
\bibitem{5} Matarrese, S., Lucchin, F., Moscardini, L. and Saez, D. 1992,
MNRAS {\bf 259}, 437.
\bibitem{6} Sathiaprakash, B.S., Sahni, V., Munshi, D., Pogosyan, D. and
  Mellot, A.L., 1995, MNRAS {\bf 275}, 463.
\bibitem{7} Mariano, P.M. 2002, Advances in Applied Mechanics {\bf 38}, 1.
\bibitem{8} Newell, A.C. 1974, Lect. Appl. Math. {\bf 15}, 157.
\bibitem{9} Aranson, I. and Kramer, L. 2002, \rmp {\bf 74}, 99.
\bibitem{10} V\'an, P. 2003, Annalen der Physik {\bf 12}, 146.
\bibitem{11} Harvey, R.J. 1966, Phys. Rev. {\bf 152}, 1115.
\bibitem{12} Spiegel, E.A. 1980, Physica {\bf 1D}, 236.
\bibitem{13} Widrow, L.M. and Kaiser, N. 1993, ApJ {\bf 416}, L71.
\bibitem{14} Coles, P. and Spencer, K. 2003, MNRAS {\bf 342}, 176.
\bibitem{15} Madelung, E. 1926, Z. Phys. {\bf 40},  332.
\bibitem{16} Salue\~na, C., Esipov, S. and P\"oschel, T. 1997, SPIE
{\bf 3045}, 2.
\end{thebibliography}
\end{document}